\documentclass[twocolumn,twocolappendix]{aastex63}

\shorttitle{Irradiation of the donor in an AM~CVn}
\shortauthors{Rivera Sandoval et al.}

\graphicspath{{./}{figures/}}
\begin{document}

\title{A year long superoutburst from an ultracompact white dwarf binary

reveals the importance of donor star irradiation}

\correspondingauthor{Rivera Sandoval, L.E.}
\email{liliana.rivera@ttu.edu}

\author{Rivera Sandoval, L.E.$^{1}$, Maccarone, T.J.$^{1}$, Pichardo Marcano, M.}
\affiliation{Texas Tech University, Department of Physics \& Astronomy, Box 41051, 79409, Lubbock, TX, USA\\}

\nocollaboration{2}

\begin{abstract}
SDSS J080710+485259 is the longest period outbursting ultracompact white dwarf binary. Its first ever detected superoutburst started in November of 2018 and lasted for a year, the longest detected so far for any short orbital period accreting white dwarf.  
Here we show that the superoutburst duration of SDSS J080710+485259
exceeds the $\sim2$ months viscous time of its accretion disk by a factor of about 5.
Consequently it follows neither the empirical relation nor the theoretical relation between the orbital period and the superoutburst duration for AM CVn systems.
Six months after the end of the superoutburst the binary remained 0.4 mag brighter than its quiescent level before the superoutburst. We detect a variable X-ray behavior during the post-outburst cooling phase, demonstrating changes in the mass accretion rate. We discuss how irradiation of the donor star, a scenario poorly explored so far and which ultimately can have important consequences for AM~CVns as gravitational wave sources, might explain the unusual observed features of the superoutburst.

\end{abstract}

\keywords{White dwarfs --- Accretion, accretion disks ---  Stars: individual (SDSS~J080710+485259) ---  Binaries: general --- Stars: dwarf novae}

\section{Introduction} \label{sec:intro}

AM~CVns are rare ultracompact binary systems in which white dwarfs (WDs) accrete matter from He-rich stars. They are characterized by short orbital periods ($P_{orb}$) in the range $\sim5-68$ min \citep[e.g.][]{2018ramsay,2020Green}.
SDSS~J080710+485259 (hereafter SDSS~0807) was recently discovered \citep{2018kong} as a member of the AM~CVns due to its lack of H and abundance of He in its spectrum. In 2018 SDSS~0807 brightened above its quiescent level of 20.8 mag in $G$ and began its first detected superoutburst. It became bright enough to be observed by small ground-based telescopes and thus a periodic signal of $53.3\pm0.3$~min was identified \citep{2019kupfer}. That modulation likely corresponds to the ``superhump" period ($P_{sh}$), which is observed during superoutbursts and represents a beat  between the orbit and a much longer precession period. Since $P_{sh}$ is typically a few percent longer than $P_{orb}$, it can be used as a proxy in the absence of $P_{orb}$. In fact, for many members of the AM~CVn family, only $P_{sh}$ has been determined because these binaries are too faint in quiescence ($<20$ mags) to be observed by small telescopes in short exposures \citep[e.g.][]{2018ramsay}. 
In this paper we present analysis of the light~curve of SDSS~0807 during superoutburst and discuss effects that might explain its peculiar observed characteristics. 

\section{OBSERVATIONS AND DATA ANALYSIS} \label{sec:style}

\subsection{Optical data}

We used public optical observations of SDSS~0807 taken with the \textit{Zwicky Transient Facility (ZTF)} in the $g$ and $r$ filters \citep{2019ZTF}. Data were obtained from 2018-03-27 to 2019-12-29 with a gap from 2019-05-14 to 2019-08-28 due to the object's occultation. Public data obtained with the $Gaia$ observatory \citep{2016gaia,2018gaia} in the $G$ filter were also used for the light~curve analysis. The first $Gaia$ measurement corresponds to 2016-04-08,  which together with data from 2017 and early 2018 helped to establish the full quiescent level. The last $Gaia$ data point was obtained on 2020-06-01 during the post-outburst cooling phase. The \textit{ZTF} data are calibrated to the AB magnitude system and only the observations flagged as good quality measurements were used for the analysis. 

We looked for indications of outbursts or superoubursts of SDSS~0807 in databases such as the \textit{Catalina Sky Survey}, and \textit{Pan-STARRS} over the last 10 years,  and no indications of such an event were found. The monitoring cadence of these surveys was not short enough to provide a tight constraint on the existence of these events, but a long superoutburst as the one here presented likely did not occur. 
SDSS~0807 was not observed by \textit{ASAS-SN} or \textit{DASCH}. 

\subsection{X-ray and UV data}
The X-ray (0.3--10 keV) and UV data analyzed in this paper were taken with the \textit{Neil Gehrels Swift Observatory (Swift)} on 2020-04-15 and 2020-05-28. A total of $3300$ s were obtained in both observations, which correspond to $1700$~s and $1600$~s, respectively. 
Data were reprocessed and analyzed following the standard reduction threads\footnote{\url{https://www.swift.ac.uk/analysis/xrt/} and \url{https://www.swift.ac.uk/analysis/uvot/}}, which make use of XSPEC v12.11.0 \citep{199arnaud}.
To determine the X-ray flux we have assumed an absorbed power-law model (TBabs*pegpwrlw) to fit the X-ray spectrum obtained on 2020-04-20. We set the value of the neutral H column ($N_H$) to the Galactic value towards the binary's position ($N_H=3.89\times10^{20}$ cm$^{-2}$). Given the small number of counts, we used C-statistics for the fit. 

Observations in the optical and UV were obtained with UVOT and the corresponding magnitudes and exposures times per observation are given in table \ref{table:uvotmags}.
For the UVOT measurements we used a circular region with a radius of $5''$ centered on RA$=08$:$07$:$10.33$, DEC$=+48$:$52$:$59.6$. A circular region with radius $30''$ and located in a star-free region of the image close to the target was used for the background subtraction. Given the coordinates of SDSS~0807 it is difficult or not possible to observe it with ground- or space-based telescopes from mid May to late August. But a few data points using \textit{Swift} and $Gaia$ were obtained during the post-outburst cooling period. 

\begin{table} 
\begin{center}
\begin{tabular}{ |c | c | c|  c|}
\hline
 Date & Filter & AB mag & Exp.time (s)  \\ 
 \hline 
 April 15 2020 & $V$ &   $> 20.13$ & 131 \\  %1s
`' & $B$ & $19.60 \pm 0.30$ & 132 \\ %3s
`' & $U$ & $20.48 \pm 0.31$ & 132 \\ %3s
`' & $UVW1$ &  $20.03 \pm 0.18$ & 262 \\ %3s
`'& $UVW2$ & $20.52 \pm 0.16 $ & 526\\ %3s
`' & $UVM2$ & $20.51 \pm 0.19$ & 436  \\ %3s
\hline
 May 28 2020 & $U$ &  $20.87 \pm 0.32$ & 634\\  %3s
`' & $UVW1$ &  $20.37 \pm 0.21$ & 439  \\   %3s
`'& $UVW2$ &  $20.47 \pm 0.18$ & 439 \\ %3s
\hline
\end{tabular}
\caption{UVOT magnitudes taken with the \textit{Swift} Observatory.}
\label{table:uvotmags}
\end{center}
\end{table}

\subsection{Spectral analysis}

We analyze public \textit{Sloan Digital Sky Survey (SDSS)} spectra of SDSS~0807, and for comparison purposes, data from SDSS~J141118.31+481257.6 (hereafter SDSS~1411). The latter is another long period ($P_{orb}=46$~min) AM~CVn system which was also recently identified in superoutburst for the first time, with an amplitude of $\sim7$~mags in optical \citep{2019RS}. A total of four and three spectra were analyzed for SDSS~0807 and SDSS~1411, respectively. For SDSS~0807 the data consist of two pairs of spectra taken sequentially and divided in two observations. The first one observed on 2014-10-03 and the second one on 2014-10-05. A total exposure time of 3600~s was obtained for that binary. For SDSS~1411, the observations consist of one exposure of 1000 s and two observations of 800~s each taken sequentially on 2005-03-17. All spectra cover the range $3800-9200$~\AA \ with a resolution of 1500 at $3800$~\AA \ (the blue channel), and 2500 at 9000~\AA \ (the red channel).
For each source, SDSS~0807 and SDSS~1411, the full width at half maximum (FWHM) values were obtained from Gaussian fits to individual $5875$~\AA\ He-I profiles. Continuum rectified spectra were fitted in a window of $200$~\AA, centered \mbox{on that He-I line (Appendix \ref{k2}).}

\begin{figure*}
\centering
\includegraphics[width=1.3\columnwidth, trim= 0cm 0cm 0cm 0cm]{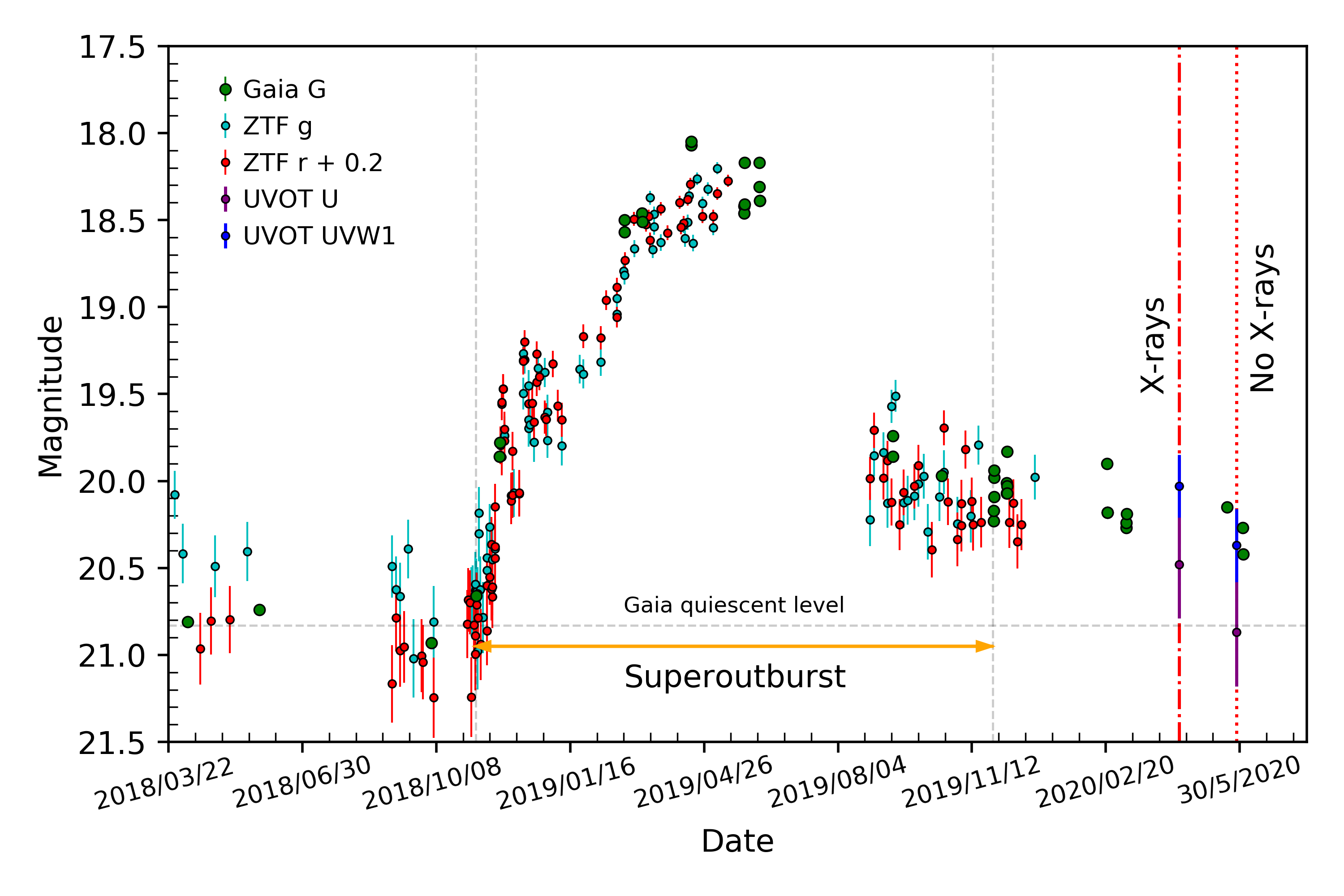}
\caption{Light~curve of SDSS~0807 with data from \textit{Gaia, ZTF} and \textit{Swift}-UVOT. The average quiescent value is marked with a horizontal line. Additional $Gaia$ points used to establish the quiescent level are not plotted. The duration of the superoutburst is indicated with an orange arrow. $ZTF$ and UVOT magnitudes are given in AB system. A constant to the magnitudes of \textit{ZTF-r} has been added to match the quiescent level of the $Gaia$ measurements. The detection ($10.1\pm 2.7 \times 10^{-3}$ counts s$^{-1}$) and non detection ($<4.5\times10^{-3}$ counts s$^{-1}$) of X-rays in the 0.3-10 keV band are also marked with red vertical lines. Lack of data during the superoutburst is due to the binary's occultation.}
\label{fig:LC}
\end{figure*}

\begin{figure*}
\centering
\includegraphics[width=1.3\columnwidth, trim= 0cm 0cm 0cm 0cm]{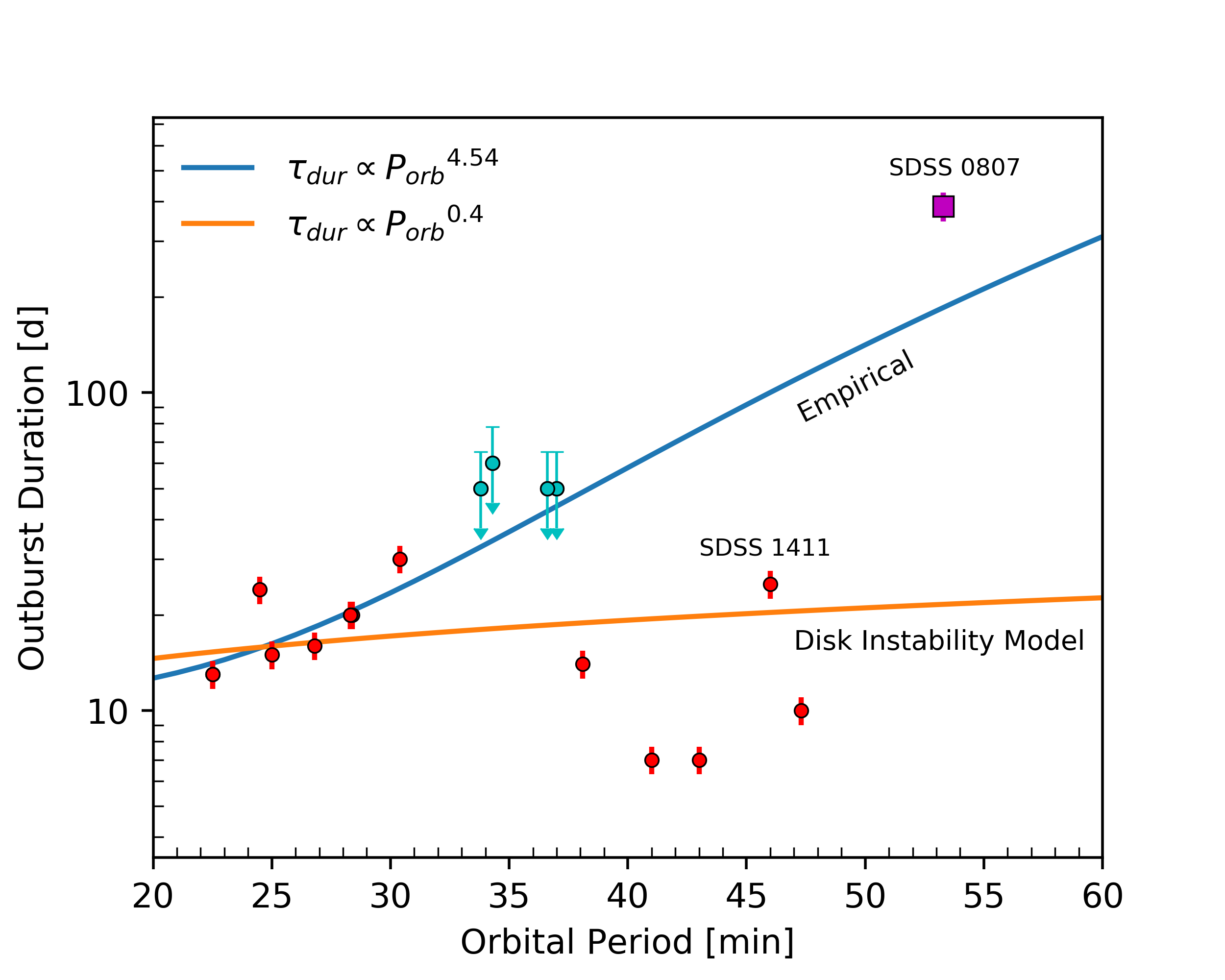}
\caption{The orbital period ($P_{orb}$) vs outburst duration ($\tau_{dur}$) relation for AM~CVns.The red circles are known AM~CVns with measured outburst duration \citep[][and references therein]{2019cannizzo}.
The cyan circles indicate $75\%$ of the upper limit values (i.e. $75\%$ probability the object is fainter than such value), which considering the remaining $25\%$ error and the values of the outburst duration at $P_{orb}<40$ min give the empirical relation \citep{2015Levitan} $\tau_{dur} \propto P_{orb}$$^{4.54}$. The relation $\tau_{dur} \propto P_{orb}$$^{0.4}$ was obtained considering the outburst duration of systems with $P_{orb}<50$ min and the DIM for He~dominated accretion disks \citep{2019cannizzo}. The observed $\tau_{dur}$ for SDSS~0807 is well above the expected value of both relations. The aforementioned relations are actually valid for superoutbursts \citep[most of these observational points and upper limits;][]{2015Levitan,2019cannizzo}, thus they are relevant for our data and discussion on SDSS~0807.
We have considered a $10\%$ error for the outburst duration for all systems that do not have an upper
limit, as done by \cite{2015Levitan}.
}
\label{fig:relation}
\end{figure*}

\section{RESULTS AND DISCUSSION}

\subsection{The superoutburst of SDSS~0807}

In Fig.\ref{fig:LC} we show the combined light~curve of SDSS~0807. 
$Gaia$ data were used as reference for the superoutburst parameters given the smaller photometric errors. The superoutburst was considered to start on 2018-11-07 when $Gaia$ measurements showed a brightness increase of $\sim0.2$ mags above the quiescent level. Previous $Gaia$ measurements (not plotted in Fig.\ref{fig:LC}) are scattered around the marked quiescent level. The peak value was given by the brightest $Gaia$ measurements. The recording of several data points, which are consistent with each other, shows that the identified magnitude value was not due to systematics. The average magnitude of the several $Gaia$ observations taken on 2019-11-28 was consistent with the one from observations on 2019-12-08 and thus, defined the end of the superoutburst. Previous measurements were above that ``stable level", suggesting that the object’s flux was still decreasing. To account for this uncertainty we have included a $10\%$ error in the duration of the superoutburst (Fig.\ref{fig:LC}).
The event lasted $\sim 390$ days, which is the longest accretion outburst ever observed for either an AM~CVn or their cousins, the H-rich accreting WDs also known as cataclysmic variables (CVs). In fact, the long duration of the event excludes the possibility of a normal outburst (which last no more than a few days). 

Superoutbursts in AM~CVns are also characterized by amplitudes of several magnitudes. Indeed, for the period of SDSS~0807 an amplitude of $\sim7$ mag is expected \citep{2015Levitan}, which contrasts with the 2.7 mags observed (Fig.\ref{fig:LC}). However, as in CVs with thick accretion disks and high inclinations where the luminosity can be reduced  by 3.5 mags when observed at $i=85^\circ$ \citep{1986warnerdisk,1987warner}, a similar situation is likely to have have occurred in SDSS~0807. We measured the FWHM of the 5875 \AA \ He-I line which is prominent in AM~CVns to estimate the radial velocity of the donor ($K_2$, see Appendix \ref{k2} for details on this calculation) and compared it to that of SDSS~1411. We found that SDSS~0807 has a larger average FWHM (and hence $K_2$) than SDSS~1411. Given its longer $P_{orb}$, the broader He-I line can only be explained if SDSS~0807 has a higher inclination angle, which, in turn, would explain the small observed amplitude of the superoutburst \citep{1986warnerdisk,1987warner}. Note that the empirical relation between $P_{orb}$ and the superoutburst amplitude found by \cite{2015Levitan} was determined for AM~CVns with $P_{orb}<40$ min, which means that validity of the extrapolation to larger periods is not well established. In fact, inaccurate predictions have already been observed for other AM~CVns systems in superoutburst such as SDSS~1411, which predicted  amplitude is 5.8 mags, but it showed a $\sim7$ mag superoutburst \citep{2019RS}. This casts doubts whether the expected amplitude of SDSS~0807 is indeed $\sim7$ mag. Also, if the WD accretor in SDSS~0807 is not a very massive one (e.g. $\sim0.6$~M$_{\odot}$), this would also affect the outburst amplitude as the accretion luminosity is directly and inversely proportional to the WD mass and radius, respectively. Furthermore, \cite{kotko2012} have also shown that a larger metallicity substantially reduces the outburst amplitudes in AM~CVns.

We also determined the rise and decline timescales of SDSS~0807 which are 60~day/mag and 115~day/mag, respectively. Both are remarkable because they are the slowest ever observed in any outbursting AM~CVn or CV system. In fact, these rate values exceed by tens of times those expected for CVs \citep{1975bailey}, even if only long-duration outbursts or superoutbursts are considered \citep{2016otu}, independently of their $P_{orb}$. The viscous time sets the slow decay of superoutbursts, and for SDSS~0807, that is expected to be $\sim2$ months (see Appendix \ref{viscous}), which clearly contrasts with our observations (Fig. \ref{fig:LC}).

From the \textit{Swift} analysis we determined that on 2020-04-15 the object was clearly detected in X-rays with a count rate of $10.1\pm 2.7 \times 10^{-3}$ counts s$^{-1}$ (or $f_X=0.40^{+0.21}_{-0.14} \times 10^{-12}$ erg cm$^{-2}$ s$^{-1}$), with a photon index $\Gamma=2.46^{+0.83}_{-0.73}$. This indicates that the X-ray spectrum of SDSS~0807 is more dominated by lower energy photons in that phase, perhaps due to a lower mass WD or a more transparent boundary layer. On the other hand, the binary was not detected in X-rays on 2020-05-28, with an upper limit on the count rate of $4.5\times10^{-3}$ counts s$^{-1}$ at the 97.5\% confidence level (corresponding to 2$\sigma$ for a one-sided tail for a Gaussian distribution). The source thus clearly faded between the April and May observations, probably by a factor more than 2. Meanwhile, measurements in the $U$ band for both dates correspond to $20.48\pm0.31$ mags and $20.87\pm 0.32$ mags, respectively. The optical decline corresponds to a reduction in flux by a factor of 1.6, but given that a substantial fraction of the optical flux in these faint states comes from the radiative cooling of the WD itself, the accretion rate likely dropped by a larger factor. These measurements are in agreement within errors with those taken by the SDSS on 2000-04-25, when the object had a magnitude in $u'$ of $20.4\pm 0.06$ mags and was fully in quiescence. On the other hand, in the UV band UVW1 ($\lambda_{c}=2600$ \AA) the binary showed a slightly decrease in magnitude (0.4 mags) between both dates, suggesting a correlation with the X-ray emission decrease. Interestingly, $Gaia$ measurements indicate that $\sim6$ months after the event finished, the binary remained 0.4 mag brighter than its original quiescent level (Fig. \ref{fig:LC}). 

Enhanced emission in X-rays and UV at least 2 months after the end of an outburst has previously been observed in other AM~CVns \citep{2019RS}, and the UV has been explained as heat released by the accreting WD, which was deposited on it during the intense accretion event (the superoutburst). The X-ray emission is explained as residual accretion, as the temperature of the accreting WD after the end of the superoutburst is not high enough to release X-rays \citep{2006godon}. The \textit{Swift} X-ray detection then shows that the binary was still accreting on 2020-04-15. On the other hand, the X-ray non-detection of SDSS~0807 on 2020-05-28 does not necessarily mean that the object reached its quiescent level, but it demonstrates that abrupt changes in the mass accretion rate occurred between 140-180 days after the end of the superoutburst. Note that the emission in the bluest UVOT band UVW2 ($\lambda_{c}=1928$~\AA) was consistent within errors during both UVOT observations and there was no flux increase in that band on 2020-05-28, when the binary was not detected in X-rays. This argues against the possibility of the boundary layer becoming optically thick (and thus reducing the X-ray emission) due to an episode of large accretion, as has been observed in CVs and AM~CVns during (or near to) the outburst peak \citep[e.g.][]{2003wheatley, 2012ramsay_sup,2019RS}.

\subsection{Irradiation of the donor}

A commonly invoked scenario to explain the outbursts and superoutbursts in AM~CVns is the disk instability model  \citep[DIM; e.g.][]{1983Smak,2008Lasota,kotko2012,2015cannizzo,2019cannizzo}, given that it reproduces the relations between $P_{orb}$ and several outburst observables such as their duration ($\tau_{dur}$), the recurrence time and the mass transfer rate relatively well  \citep{2015cannizzo,2019cannizzo}. It also predicts the threshold period for stable versus unstable accretion disks \citep{kotko2012}. That model is an extension of the model applied to CVs, with the difference that the disks in AM~CVns are dominated by He instead than by H, and their sizes are much smaller. The model assumes that the mass transfer rate from the donor is constant and the value of the accretion rate onto the accreting WD determines the stability of the disk and therefore the existence of outbursts/superoutbursts. In the DIM for AM~CVns the duration of the superoutburst increases with $P_{orb}$ and follows the relation $\tau_{dur}\propto P_{orb}$$^{0.4}$ \citep{2019cannizzo}, implying that SDSS~0807 should have had a superoutburst with a duration of $22$ days, which contrasts with the $\sim390$ days observed. The observations here presented argue then against that model (Fig.\ref{fig:relation}). Comparison between the superoutburst duration and the empirical relation derived by \cite{2015Levitan} is not well-motivated as that relation was derived for systems with $P_{orb}<40$~min and for more than one detected superoutburst. Thus, we limit our discussion regarding the superoutburst duration to the DIM (Fig.\ref{fig:relation}).

A poorly explored effect so far is the importance of irradiation of the donor by the accretor and its disk, which can enhance the mass transfer rates \citep{1995Warner, 1997hameury,2007Deloye,kotko2012, 2015warner}. Given that the orbits in AM~CVns have very short periods and the donors can be quite cold, this effect might be important in at least a fraction of the AM~CVn systems.  Irradiation could, in fact, explain the long superoutbursts such as in SDSS~0807, with longer rise and decline times due to changes in the mass transfer rate. Indeed, indications of changes in the mass transfer rates of other outbursting AM~CVns have been observed \citep{2000patterson}, as have anomalously long outbursts of a few months in duration \citep{2012Ramsay,2012shears}, suggesting that irradiation on the companion can be important and even relatively common. The extreme superoutburst characteristics of SDSS~0807 are the clearest and most extreme evidence of the action of this mechanism in AM~CVns.

In fact, one can expect a significant effect on the structure of the donor star due to irradiation if the rate at which it absorbs heat from the accretor's radiative flux is comparable to the rate at which its internal energy is radiated.  If we assume an initial mass of 0.2~M$_\odot$ and an initial temperature of $10^8$ K for the donor WD (both likely higher than the typical initial values for the donors in AM~CVn systems), and then follow the cooling of the donor star \citep{1952Mestel}, taking into account that mass is lost during the evolution, then at $P_{orb}\sim53$ min, we expect the luminosity of the donor to be about $3.8\times10^{29}$ erg s$^{-1}$ and the temperature to be 2600 K. Using the same approach, we find that we expect $T=1700$~K for a 68 min orbital period, in reasonable agreement with the recent discovery of a $T \approx$ 1850 K excess in SDSS~J1505+0659 \citep{2020Green}. The donor will occupy about 0.0025 of the solid angle seen from the source.  Thus, if the superoutburst luminosity is $\gtrsim 10^{32}$ erg s$^{-1}$, the WD donor should heat substantially due to the outburst of the accretion disk, given that the albedo of a He-star absorbing UV light should be low.  Assuming a distance of at least 1~kpc (reasonable, given that the parallax to SDSS~0807 is not yet well constrained), the outburst does, indeed, reach this luminosity. 
The evolution of AM~CVns indicates that they move from a short $P_{orb}$ to a larger one as they evolve, with the donors significantly reducing their mass transfer rates and cooling as the orbit widens \citep{2005nelemans}. Thus, the long period of SDSS~0807 points towards an old, cold and low-mass donor, which  would be easy to heat during a superoutburst.These simple calculations show that the donor is susceptible to having irradiation affect it, but a more detailed quantification of the effects of donor irradiation is beyond the scope of this paper given the complexity of the problem. However, \cite{kotko2012} have shown that mass-transfer enhancement does in fact substantially increases the duration of the outbursts in AM~CVns besides affecting the shape of the light~curves. Thus, the results of SDSS~0807 will certainly place constraints for a more detailed modeling of outbursts in these binaries.

The observations of SDSS~0807 then show that irradiation on the donor has to be considered in the evolution of AM~CVns, as it can have important consequences for these binaries as gravitational wave sources. If, for instance, enhanced mass transfer due to irradiation makes AM~CVns evolve more quickly towards longer periods than currently predicted by models (which usually neglect this mechanism), then the number of ultracompact systems expected to be detected by \textit{LISA} \citep{2004nelemans,2011marsh,2012nissanke} will likely be smaller than currently thought; the \textit{LISA} noise curve's shape will change as well, although this is likely to be dominated by detached double WDs. It is also possible that the abrupt changes in the mass transfer rates make the donor unstable, and therefore after the superoutburst (when the mass transfer rate will substantially drop), it will need to re-adjust its structure. This will likely make the recurrence times much longer than currently expected \citep[either empirically or with the DIM;][]{2015Levitan,2019cannizzo}, therefore reducing the detection rate of these systems through their outbursts by all sky variability surveys, directly influencing the density estimates. Detail modeling and further investigations on the accreting behavior of AM~CVns is required in order to quantitatively assess how common this mechanism is on these binaries. Models will also help to investigate what are the causes that influence the donor's irradiation, among which could be the precession
of a tilted/warped disk \citep[e.g.][and references therein]{kotko2012} and perhaps also the metallicity due to the type of donor (and hence the formation channel).
Furthermore, to find these systems in deep quiescence new techniques should be developed. 

\acknowledgments

The authors thank the anonymous referee for his/her comments which improved this manuscript. 
This work has made use of data from the European Space Agency (ESA) mission
{\it Gaia} (\url{https://www.cosmos.esa.int/gaia}), processed by the {\it Gaia}
Data Processing and Analysis Consortium (DPAC,
\url{https://www.cosmos.esa.int/web/gaia/dpac/consortium}). Funding for the DPAC
has been provided by national institutions, in particular the institutions
participating in the {\it Gaia} Multilateral Agreement. We acknowledge the Photometric Science Alerts Team (\url{http://gsaweb.ast.cam.ac.uk/alerts}). Based on observations obtained with the Samuel Oschin 48-inch Telescope at the Palomar Observatory as part of the Zwicky Transient Facility project. ZTF is supported by the National Science Foundation under Grant No. AST-1440341 and a collaboration including Caltech, IPAC, the Weizmann Institute for Science, the Oskar Klein Center at Stockholm University, the University of Maryland, the University of Washington, Deutsches Elektronen-Synchrotron and Humboldt University, Los Alamos National Laboratories, the TANGO Consortium of Taiwan, the University of Wisconsin at Milwaukee, and Lawrence Berkeley National Laboratories. Operations are conducted by COO, IPAC, and UW. 
The authors also acknowledge the \textit{Swift} team for scheduling the target of opportunity requests, the \textit{SDSS} and \textit{vizier} data bases for providing part of the data presented in this manuscript.

\bibliography{biblio.bib}
\bibliographystyle{aasjournal}

\appendix
\section{Qualitative estimation of $K_2$}
\label{k2}

We used the measured FWHM and the periods to estimate the difference in the inclination of the two sources, SDSS~0807 and SDSS~1411. We assumed that the FWHM of the 5875 \AA\ He-I line is correlated with $K_2$ in an 
analogous way to that found by \cite{Casares2015} for black holes, where
the FWHM is related to the radial velocity semi-amplitude of the donor star ($K_2$) as

$$ \left ( \frac{FWHM}{2} \right )^2 = \frac{G M_1}{R_d} \sin^2 i  $$

and 

$$K_2^2 = \frac{G M_1^2}{a (M_1+ M_2)} \sin ^2 i $$

\noindent where $R_d$ is the radius of the accretion disk, $M_1$ is the mass of the accretor, $M_2$ is the mass of the companion star, and $i$ is the binary inclination angle. Assuming that the disk radius, $R_d = \alpha R_{L1}$ where $\alpha < 1$ and $R_{L1}$ is the Roche lobe radius and can be computed from the Eggleton’s relation \citep{Eggleton83}:

$$\frac{R_L}{a} = \frac{0.49 q ^{-2/3}}{0.6 q ^{-2/3} + \ln (1+ q^{-1/3})}$$

We obtain $K_2$ as a function of $\alpha$ and the mass ratio $q=M_2/M_1$, i.e.

$$\frac{K_2}{FWHM}  = \frac{\sqrt{\alpha f(q)}}{2}  $$

where 

$$f(q) =  \frac{0.49 (1+q)^{-1} }{0.6 + q ^{2/3}  \ln (1+ q^{-1/3})}$$

For AM~CVns there are 17 systems with reported mass ratios ranging from 0.014 to 0.01 \citep{Green2018q,2020Green}. This means that $K_2$ depends very modestly of $q$ and for this range of mass ratios the term $\sqrt{f(q)}$ varies between 0.67 to 0.8. The constant number $\alpha$ has been reported for the eclipsing AM~CVn Gaia14aae \citep{Green2018gaia14aee} to be 0.8. 

The relationship between $K_2$ and the FWHM for black hole X-ray binaries \citep{Casares2015} has been found to be:

$$K_2= 0.233\cdot FWHM$$

\begin{figure*}
\centering
\begin{minipage}{\columnwidth}
  \centering
  \includegraphics[width=\columnwidth]{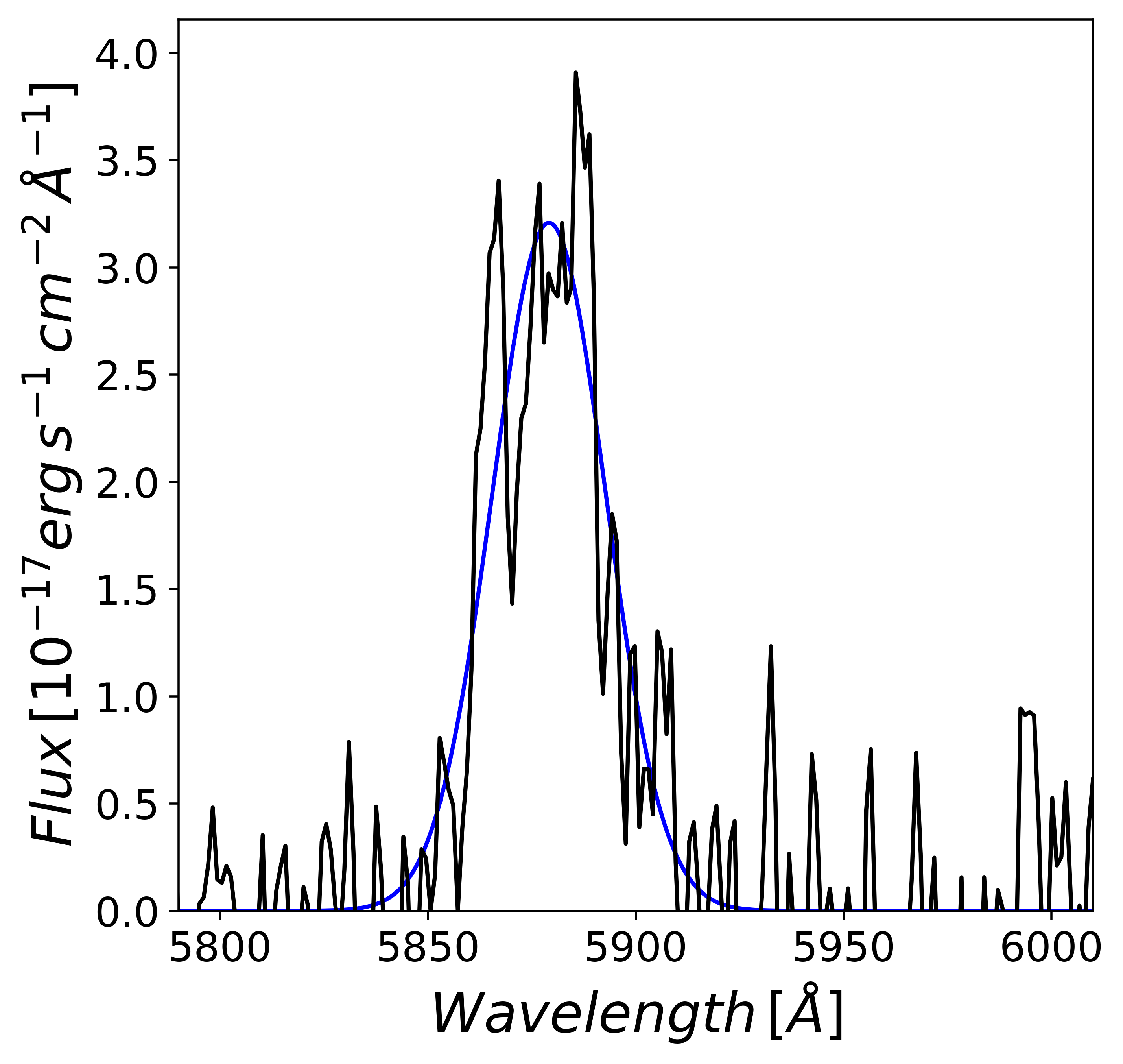}
\end{minipage}%
\begin{minipage}{\columnwidth}
  \centering
  \includegraphics[width=\columnwidth]{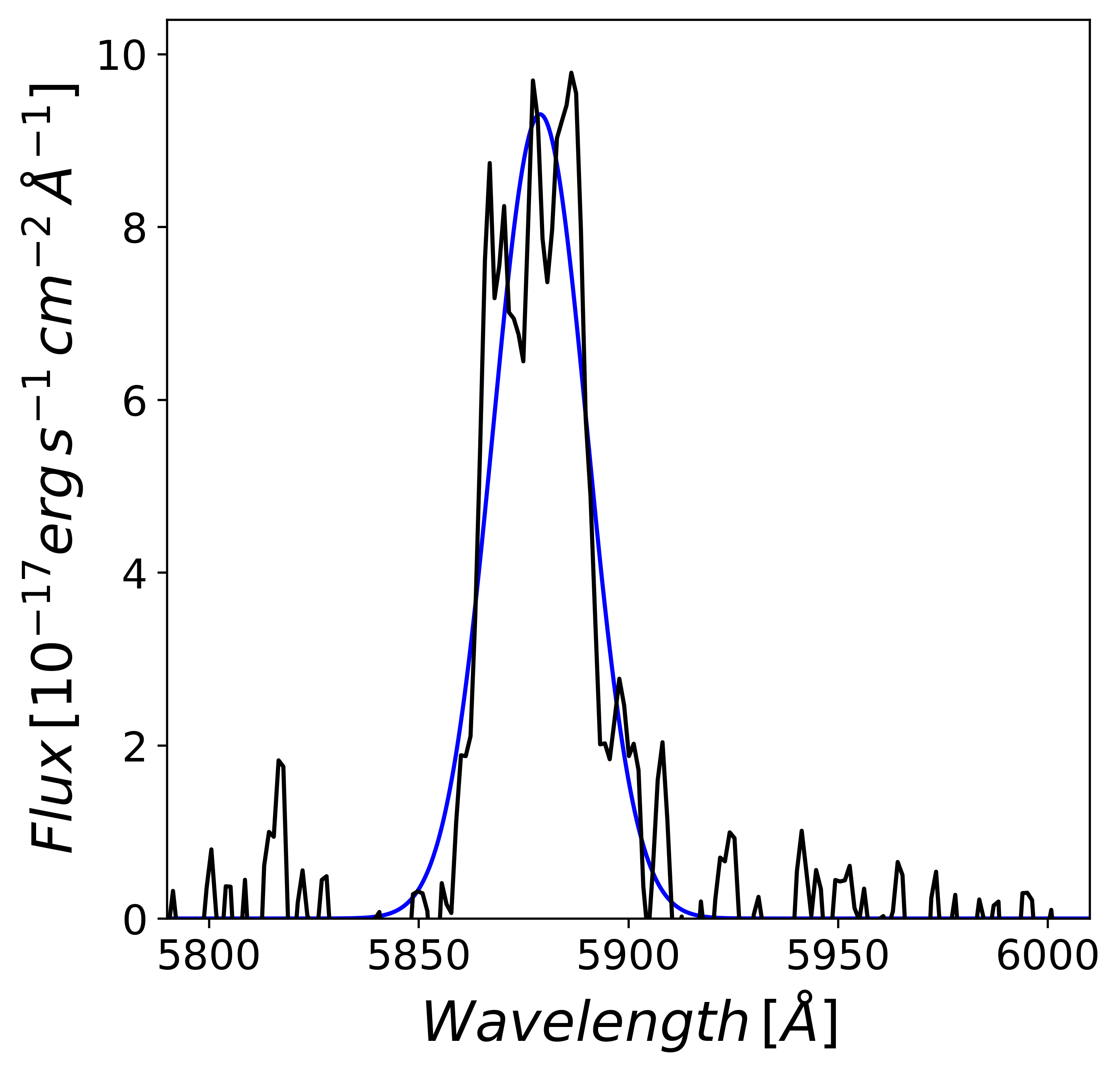}
\end{minipage}
\caption{SDSS average spectra of the AM~CVn systems SDSS~0807 (left) and SDSS~1411 (right) in the range 5800-6000 \AA. The Gaussian fit to the 5875 \AA\ He-I emission line is shown in blue color. The average FWHM are FWHM$_{0807} = 1670 \pm 336 $~km/s and FWHM$_{1411} = 1350 \pm 85 $~km/s, respectively.}
\label{fig:fwhm}
\end{figure*}

Similarly for AM~CVns, we can then expect $K_2$ to be proportional to the FWHM, which together with $P_{orb}$, can be used to determine their relative inclinations. SDSS~0807 has a $P_{orb}$ close to the likely measured $P_{sh}$ of 53.3 min, and SDSS~1411 has a $P_{orb}=46$ min. The average FWHM using all the available spectra are FWHM$_{0807} = 1670 \pm 336 $ km/s and FWHM$_{1411} = 1350 \pm 85 $ km/s, respectively (Fig. \ref{fig:fwhm}). Since $K_2$ is inversely proportional to $P_{orb}^{1/3}$, 
for SDSS~0807, we would then expect a smaller FWHM than that of SDSS~1411. 
However, the observations indicate the contrary, suggesting that the larger FWHM of SDSS~0807 is then due to a larger inclination angle when compared to SDSS~1411. 
Note that here we do not give a explicit value of the constant that relates $K_2$ and the FWMH for AM~CVns. At the present time there is insufficient data to calibrate these relations for AM CVns, using the He-I line, because the ionization structure of the disk affects the overall constant of proportionality differently for each emission line. 
Also note that from the photometric data presented in this paper we can not exclude the presence of an eclipse as in the case of Gaia14aae \citep{2015campbell}, an AM~CVn with $P_{orb}\sim50$~min which had a larger outburst amplitude (5~mags) than the one of SDSS 0807. However, \cite{2019kupfer} did not mention the identification of such characteristic in their light~curve analysis of SDSS~0807 using short-cadence observations.
The result that SDSS~0807 is significantly more edge-on than SDSS~1411 is, nonetheless robust.

\section{Viscous time}
\label{viscous}

We have estimated the viscous time of SDSS~0807 by assuming a primary mass of $m_1=0.6$~M$_\odot$ and a donor mass of $m_2=0.01$~M$_\odot$. We considered $\dot M \gtrsim 7\times 10^{-11}$~M$_\odot \ yr^{-1}$ which corresponds to at least 10 times (due to the superoutburst) the quiescent value of SDSS J1208, an AM~CVn with a similar orbital period  \citep[$P_{orb}=53$ min for SDSS J1208;][]{2018ramsay} to that of SDSS~0807. We also considered $\alpha=0.1$ and the following formula \citep{frank2002accretion}:

$$t_{visc}\sim 3\times 10^5 \alpha^{-4/5} \dot M_{16}^{-3/10} m_1^{1/4} R_{10}^{5/4} \ s $$

\noindent where $R_{10}=R/10^{10}$ cm is the radius of the disk, considered here to be the tidal radius $R=0.8a$, and $\dot M_{16}=\dot M/10^{16}$ g/s. We obtained a value $t_{visc}\sim66$ days.



\end{document}